\documentclass{raa}
\usepackage{graphicx,times}
\usepackage{natbib}
\usepackage{lscape}
\usepackage{longtable}
\usepackage{amssymb,amsmath}

\begin{document}

	\title{Infrared Photometric Properties of Inner and Outer Parts of HII regions}

	\volnopage{ {\bf 20XX} Vol.\ {\bf X} No. {\bf XX}, 000--000}
	\setcounter{page}{1}

	\author{A.P. Topchieva\inst{1}, V. V. Akimkin\inst{1}, G. V. Smirnov-Pinchukov\inst{2}}

	\institute{Institute of Astronomy, Russian Academy of Sciences, Moscow 119017, Russia; {\it ATopchieva@inasan.ru}\\
		\and
		Max Planck Institute for Astronomy, K{\"o}nigstuhl 17, D-69117 Heidelberg, Germany\\
	}
	\vs \no
	{\small Received ...; accepted ...}

	\abstract{The fact that infrared ring nebulae (IRRNs) are frequently associated with HII regions, provides us the opportunity to study dust at the interface between ionized and neutral gas. In this paper, we analyze the infrared radiation in the range from 8 to 500~$\mu$m in outer and inner parts of 32 IRRNs showing round shape. We aim to determine the morphology of these objects and possible dust evolution processes on the base of the comparison of IR radiation towards the ionized and neutral regions. We calculate six slopes between adjacent wavelengths in their spectral energy distributions to trace the difference in the physical conditions inside and outside ionized regions. Using the data on these 32 objects we show that their morphology is likely 3D spherical rather than 2D plane-like. The slope between 70 and 160~$\mu$m is the most appropriate tracer of the dust temperature in the outer envelope. The larger 8-to-24~$\mu$m intensity ratio is associated with smaller intensities at mid-IR indicating that the PAHs may indeed be generated due to larger grain destruction. These data are important for the subsequent theoretical modeling and determining the dust evolution in HII regions and their envelopes.
	}

	\authorrunning{A.~P.~Topchieva et al. }            
	\titlerunning{Infrared Photometric Properties of HII regions}  
	\maketitle

	%
	\section{Introduction}           
	\label{sect:intro}

	Thanks to infrared (IR) observations by space telescopes {\it Spitzer}, WISE, Akari and {\it Herschel}, more than 8\,000 HII regions were discovered~\citep{2006ApJ...649..759C, 2007ApJ...670..428C, Anderson_2012, Anderson_2014, Bufano}, which look like full or open infrared ring nebulae (IRRN). Most of these objects are identified as HII regions around massive stars of O--B type~\citep{2006ApJ...649..759C, 2007ApJ...670..428C} or as Wolf--Rayet stars~\citep{Gvaramadze_2010}. A specific feature of IRRNs is the presence of an outer ring visible in the near, middle and far IR from 8 to 500~$\mu$m, while emission from the inside of the ring is observed only in the middle IR within 24--70~$\mu$m, and is significantly suppressed in the near and far IR. As emission at different wavelengths can be generated mostly by dust grains with a specific size, suppressed 8~$\mu$m and far-IR emission from the ionized region can be attributed to a non-uniform distribution of dust particles in the HII regions caused by photo-destruction, radiation pressure or stellar wind~\citep{Mathis, Kruegel,2013ARep...57..573P, 2016A&A...586A.114M, Akimkin_17}.

	Statistical analysis of HII regions (see e.g.~\citet{Anderson_2012, Khramtsova_2013, Anderson_2014, Topchieva_2017, 2016A&A...586A.114M}) is a powerful tool to advance a further interpretation of observational data and to relate them to the results of numerical modeling. This is important as there are still some key questions that lack definite answers. The dominant mechanism of removing dust from HII regions is among them. In~\citet{Akimkin_17} we presented a counter-intuitive theoretical prediction that the radiative drift of dust is less efficient around stars with higher effective temperatures. Indeed, more energetic stars cause stronger radiation pressure on dust, but they also charge the grains more, which increases the dynamical coupling of dust to the plasma. The combined action of radiation pressure and Coulomb drag force leads to less effective expulsion of dust from stars with higher effective temperatures. Photo-destruction is, on the contrary, likely more effective in more energetic environments. Thus, one may expect smaller amounts of dust around hotter stars if photo-destruction dominates over radiation pressure in dust removal from the ionized region, and vice versa. The central ionizing stars are not readily identified, which hampers the determination of how the amount of dust in ionized regions depends on stellar effective temperatures.  With other things being equal, the dust temperature can potentially serve as an indicator of stellar energetics.

	In previous works we presented a multi-wavelength catalog~\citep{2017ARep...61.1015T, Topchieva}  providing total fluxes for 99 IRRNs at eight wavelengths. Qualitatively, particles of different sizes can be traced through these wavelengths. Emission at $8\,\mu$m is mostly generated by tiny polycyclic aromatic hydrocarbons (PAHs),  24~$\mu$m emission comes mostly from very small grains (VSGs) with size $\sim50$\AA, whereas larger grains ($<10^{-5}$~cm) contribute in 70--500~$\mu$m wavelength range.

	In this paper, we aim to use the dust emission towards the ionized and neutral parts of HII regions to differentiate between their 2D toroidal vs 3D spherical morphology and to constrain the dust removal mechanisms. To do this we update our catalog~\citep{Topchieva} with new information on radiation fluxes from the inner and outer parts of IRRNs associated with HII regions. In Section 2 we describe the data processing. In Section 3 we analyze the spectral differences between the inner and outer parts of IRRNs. In Section 4, we summarize the obtained results.

	\section{Data processing}
	\label{sect:Obs}
	In this work, we used the data from the previously developed catalog~\citep{2017ARep...61.1015T}, which includes 99 IRRNs associated with HII regions. The work is based on the 8~$\mu$m survey obtained on the GLIMPS camera\footnote{Galactic Legacy Infrared Surplane Extraordinaire}~\citep{2004ApJS..154...25R, 2004ApJS..154...10F}, and on 24~$\mu$m MIPSGAL survey\footnote{24 and 70 Micron Survey of the Inner Galactic Disk with MIPS}~\citep{2003PASP..115..953B, 2009PASP..121...76C} from Spitzer telescope. We also used HiGAL\footnote{https://tools.asdc.asi.it/HiGAL.jsp} data from the Herschel Space Telescope at wavelengths of 70, 160, 250, 350, and 500~$\mu$m. Point sources were removed from the images using automatic cleaning procedure described in \citet{2017ARep...61.1015T}. The search and removal of point sources are carried out in three subsequent stages depending on the brightness of the stars. In our previous study, we fitted 99 IR--ring nebulae with ellipses to determine the position of the center of the nebula, its size, the degree of asymmetry, and the position angle. Based on these results, we selected 32 objects, for which the eccentricity of the fitted ellipses does not exceed 0.6 at the wavelength of 8~$\mu$m, and the semi-major axis exceeds $20''$. These `perfect' IR ring nebulae are further analyzed in this study. The size and position of the center of the aperture superimposed on the source are also determined from the 8~$\mu$m images, as well as the values of the major ($a$) and minor ($b$) axes~\citep{2017ARep...61.1015T}. We used 8~$\mu$m data to find the size of the object as the radiation from the outer ring at 8~$\mu$m is present for all object in the sample and the boundaries of the object can be defined clearly.

\begin{figure}[h]
		\centering \includegraphics[width=0.4\textwidth]{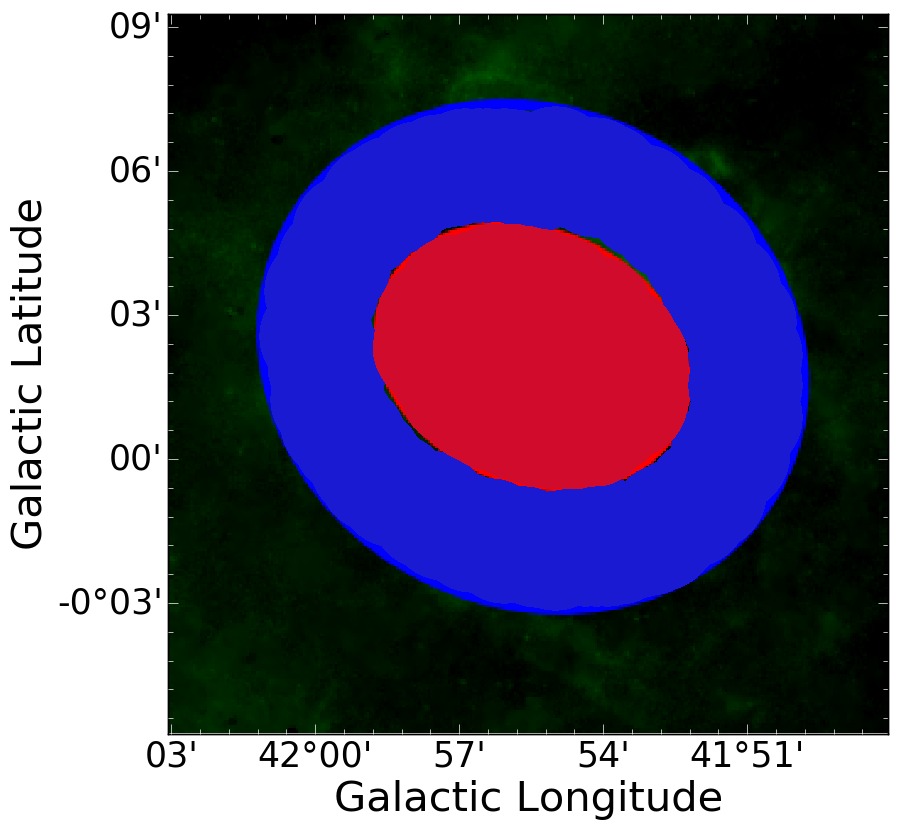}
		\includegraphics[width=0.4\textwidth]{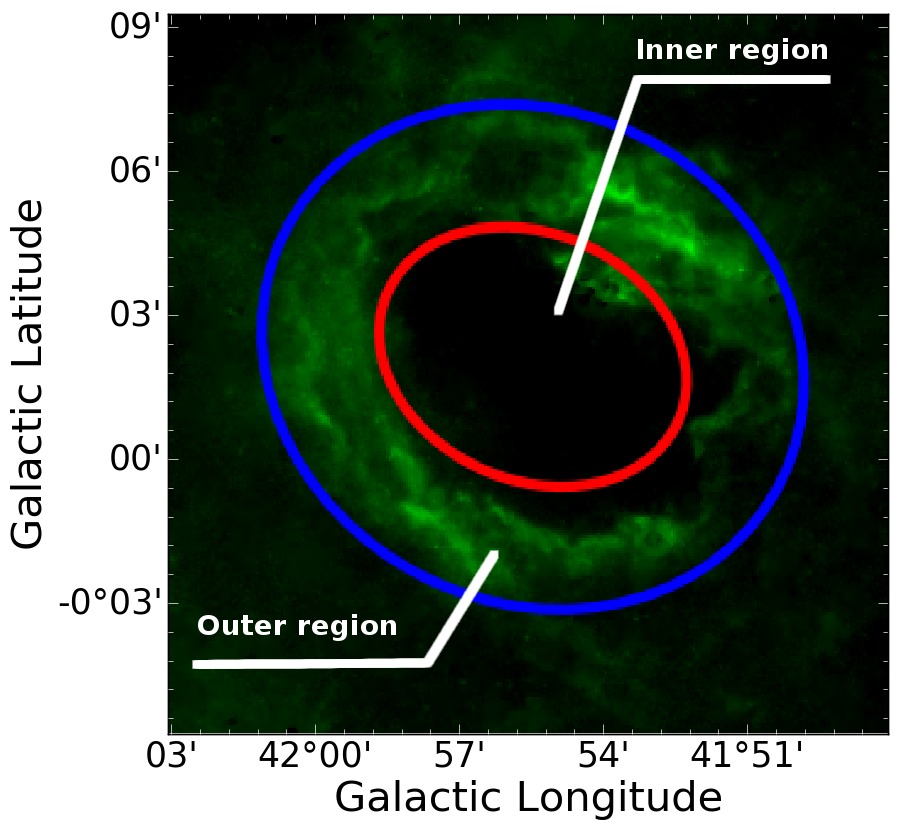}
		\caption{An example of apertures for calculating the flux from the outer and the inner regions for N~80. The green color on the second panel represents the 8~$\mu$m emission.}
		\label{fig1}
	\end{figure}

	For the selected 32 objects we define fluxes at seven wavelengths ranging from 8 to 500~$\mu$m for both the shoveled envelope and for the inner ionized zone. To define the apertures for the flux derivation in a standardized way we use the following procedure. As the typical full width of the outer rings in near-IR is $20-50$\% of its radius, we define the inner aperture as an ellipse with major and minor axes $a_{\rm inn}=0.7a$ and $b_{\rm inn}=0.7b$. This choice allows to exclude most of the envelope side radiation, but surely not the radiation coming from the possible front and back walls in case of 3D morphology. The outer aperture lies between the inner aperture and encompassing ellipse with $a_{\rm out}=1.3a$ and $b_{\rm out}=1.3b$. The example of such choice of apertures for N~80 is presented in~ Fig.~\ref{fig1}.
	We intentionally do not subtract the 'background' radiation as its estimate using the radiation far from the object in the image plane may introduce wavelength dependent bias detrimental for the purpose of the study. This is caused by small but noticeable differences in the radiation spectrum of the matter surrounding the object from different sides.

	\section{Comparison of spectral characteristics of inner and outer regions}
	\subsection{Spectral energy distributions}

\begin{figure}[t]
		\centering \includegraphics[width=0.45\textwidth]{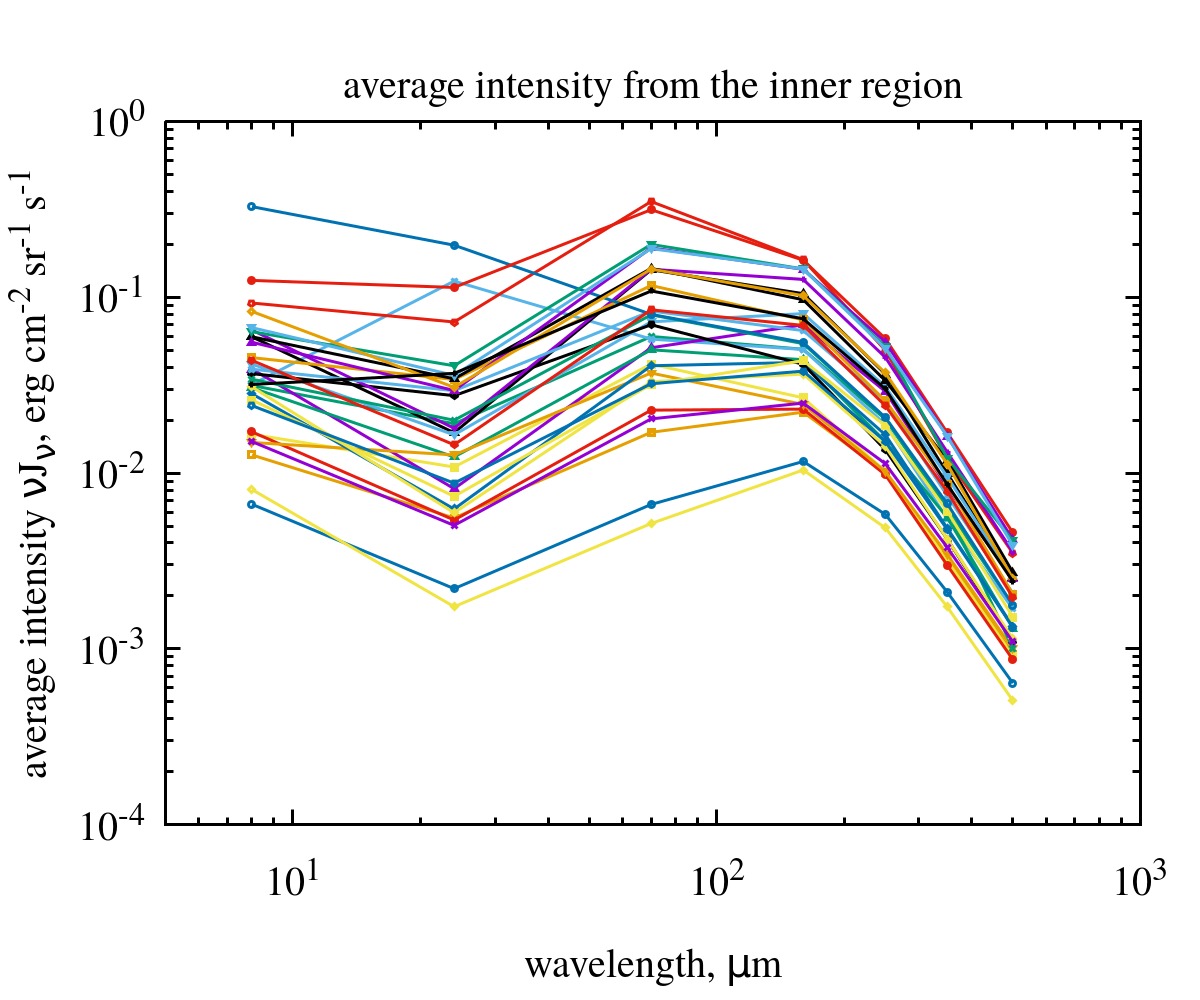}
		\includegraphics[width=0.45\textwidth]{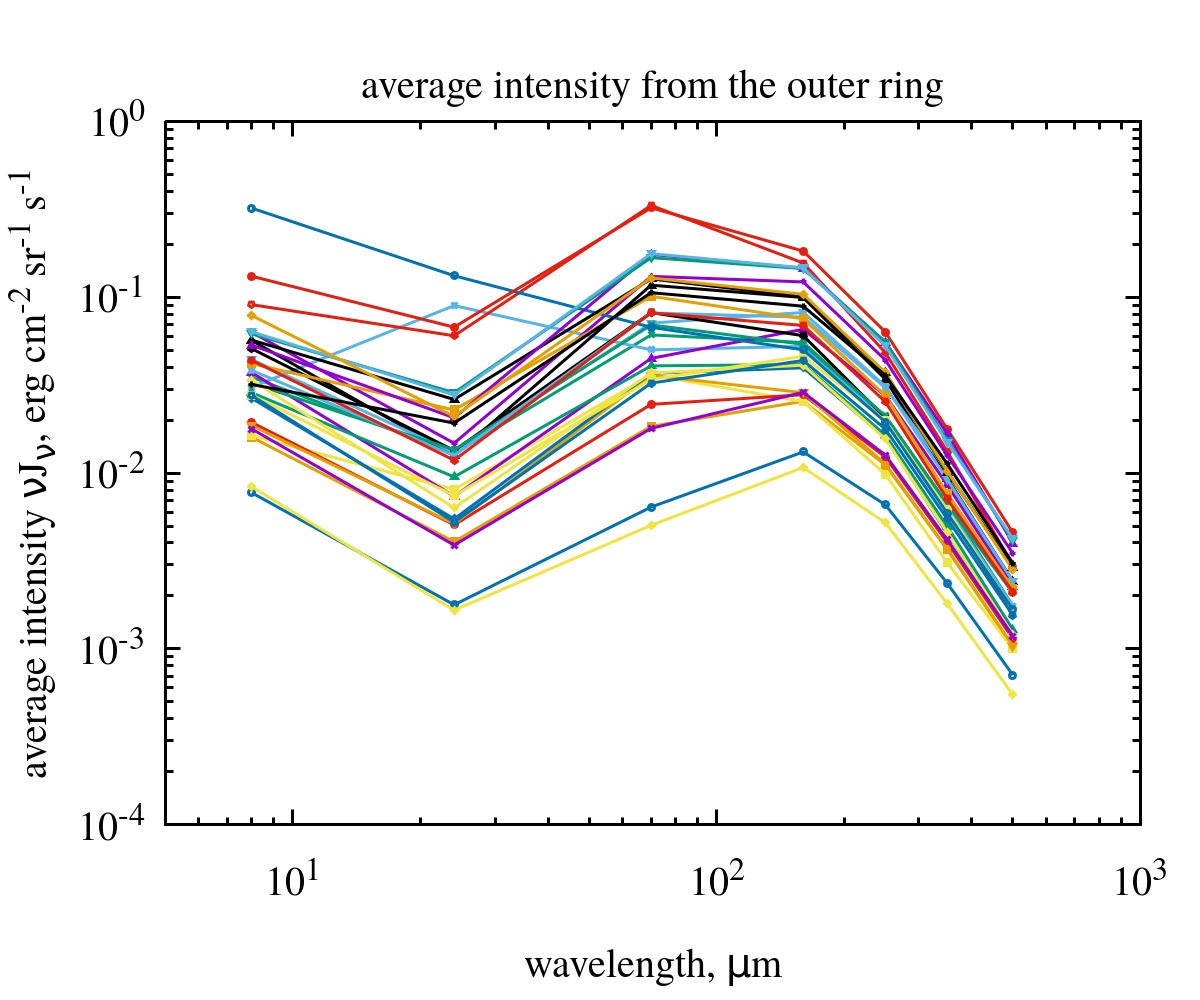}
		\caption{Spectral energy distributions from inner (left) and outer (right) parts of 32 IR--ring nebulae.}
		\label{fig2}
	\end{figure}

	For all wavelengths (8, 24, 70, 160, 250, 350 and 500~$\mu$m) we used the same aperture, which is determined for each object individually, as described above. The fluxes from the inner and outer parts of the objects are presented in Table~\ref{tbl-innflux} and \ref{tbl-outflux} as well as the solid angles of the apertures. By dividing the fluxes by corresponding solid angles one can get the average intensities within inner and outer regions. This allows presenting the spectral energy distribution (SED) in a unified way for all 32 objects. The SEDs for inner and outer regions are presented in Fig.~\ref{fig2}.
	The characteristic feature of these SEDs is the presence of two components, the thermal emission from cold and large dust grains peaking at $\sim 100\mu$m and the near-IR stochastically generated emission from PAHs and very small grains.

	{\scriptsize
		\begin{longtable}{lccccccccc}

			\caption{Fluxes and solid angles for the inner regions of 32 IR ring nebulae. Objects are taken from $^1$~\citet{2012MNRAS.424.2442S}, $^2$~\citet{2006ApJ...649..759C}, $^3$~\citet{2003yCat.5114....0E}.}\label{tbl-innflux}\\ \hline  \hline
			Object & ($l_{\rm gal}^{\circ}$; $b_{\rm gal}^{\circ}$) & $F_{8}$,Jy & $F_{24}$,Jy & $F_{70}$,Jy & $F_{160}$,Jy & $F_{250}$,Jy & $F_{350}$,Jy & $F_{500}$,Jy & $\Omega,10^{-8}$sr \\ \hline
			\endfirsthead \hline
			\multicolumn{10}{|c|}{\scriptsize\slshape(Continuation)} \\ \hline
			Object & $l_{\rm gal}^{\circ}$; $b_{\rm gal}^{\circ}$ & $F_{8}$,Jy & $F_{24}$,Jy & $F_{70}$,Jy & $F_{160}$,Jy & $F_{250}$,Jy & $F_{350}$,Jy & $F_{500}$,Jy  & $\Omega,10^{-8}$sr \\ \hline
			\endhead \hline
			\multicolumn{10}{|c|}{\scriptsize\slshape To be continued} \\ \hline
			\endfoot \hline
			\endlastfoot
			S123$^{2}$                  & (312.97; --0.43)  &43.1   &46.8    &502.4   &1354.0  &840.2  &372.2   &148.1 &164  \\
			S167$^{2}$                  & (301.62; --0.34)  &90.5   &89.9    &790.4   &3173.0  &2476.0 &1255.0  &543.2 &1250   \\
			CN111$^{2}$                 & (8.31; --0.08)    &34.8   &41.4    &531.0   &1351.0  &812.1  &348.5   &132.1 &76.9   \\
			N67$^{2}$                   & (35.54; 0.01)     &6.8    &12.3    &113.3   &218.9   &136.9  &59.5    &23.8  &19.9   \\
			N90$^{2}$                   & (43.77; 0.06)     &11.9   &15.6    &140.5   &417.8   &295.3  &136.4   &54.9  &86.5   \\
			N96$^{2}$                   & (46.94; 0.37)     &0.9    &1.8     &20.3    &29.8    &17.7   &7.7     &3.1   &5.12   \\
			N98$^{2}$                   & (47.02; 0.21)     &13.1   &12.4    &152.6   &353.5   &233.8  &100.1   &41.5  &70.3   \\
			N102$^{2}$                  & (49.69; --0.16)   &19.6   &49.8    &425.3   &644.2   &418.3  &197.5   &82.0  &121   \\
			N121$^{2}$                  & (55.44; 0.88)     &0.9    &0.6     &5.2     &24.0    &17.6   &8.7     &3.6   &10.6   \\
			N4$^{2}$                    & (11.89; 0.74)    &51.4   &115.4   &852.3   &1161.0  &592.2  &257.6   &95.9  &128   \\
			N20$^{2}$                   & (17.91; --0.68)   &10.3   &8.7     &110.8   &346.2   &230.3  &105.2   &37.6  &36.6   \\
			N14$^{2}$                   & (14.00; --0.13)   &66.8   &182.7   &1475.0  &1741.0  &973.4  &398.0   &153.3 &49.1   \\
			N23$^{2}$                   & (18.67; --0.23)   &3.6    &6.1     &76.6    &116.1   &62.8   &30.9    &10.2  &5.51   \\
			N32$^{2}$                   & (23.90; 0.07)     &3.2    &5.1     &96.5    &166.8   &100.6  &41.0    &15.1  &5.30   \\
			N33$^{2}$                   & (24.21; --0.04)   &3.4    &6.5     &93.6    &155.1   &87.7   &28.4    &13.8  &4.93   \\
			N28$^{2}$                   & (21.35; --0.13)   &6.5    &7.2     &99.5    &160.5   &91.1   &38.7    &12.3  &7.24   \\
			N51$^{2}$                   & (29.15; --0.25)   &42.1   &24.3    &395.6   &998.0   &604.9  &251.5   &97.3  &125   \\
			N60$^{2}$                   & (33.81; --0.14)   &5.1    &5.0     &86.4    &161.1   &87.9   &40.0    &14.2  &10.7   \\
			N70$^{2}$                   & (37.75; --0.11)   &3.1    &10.8    &93.1    &148.0   &92.2   &36.9    &14.8  &9.02   \\
			N80$^{2}$                   & (41.93; 0.03)     &14.6   &14.5    &171.8   &483.0   &340.8  &159.3   &66.0  &36.2  \\
			G027.492+0.192$^{3}$        & (27.49; 0.19)     &6.9    &16.0    &226.2   &241.6   &115.1  &39.1    &16.1  &6.78   \\
			MWP1G032731+002120$^{1}$    & (32.72; 0.21)     &3.0    &35.0    &47.8    &95.4    &55.9   &22.7    &10.0  &8.70   \\
			MWP1G030250+002413$^{1}$    & (30.24; 0.24)     &6.7    &5.7     &142.0   &237.4   &129.5  &49.3    &17.9  &10.4   \\
			MWP1G017626+000493$^{1}$    & (17.62; 0.04)     &2.1    &1.3     &24.6    &76.2    &49.7   &19.7    &8.5   &5.03   \\
			MWP1G018580+003400S$^{1}$   & (18.58; 0.34)     &1.6    &1.8     &22.3    &44.9    &24.0   &12.4    &4.1   &1.91   \\
			MWP1G048422+001173$^{1}$    & (48.42; 0.11)     &2.8    &1.8     &35.9    &85.9    &51.9   &21.4    &8.4   &9.25   \\
			MWP1G024558--001329$^{1}$   & (24.55; --0.13)   &13.5   &11.1    &261.1   &522.4   &294.4  &119.1   &45.9  &19.0   \\
			MWP1G037196--004296$^{1}$   & (37.19; --0.42)   &1.9    &3.4     &39.6    &62.1    &36.7   &15.5    &3.5   &5.24   \\
			MWP1G032057+000783$^{1}$    & (32.05; 0.07)     &7.6    &17.2    &142.9   &253.0   &157.3  &63.5    &25.2  &18.0   \\
			MWP1G018743+002521$^{1}$    & (18.74; 0.25)     &3.4    &7.6     &76.4    &111.3   &60.2   &28.3    &9.5   &6.89   \\
			MWP1G037349+006876$^{1}$    & (37.35; 0.68)     &44.5   &80.6    &95.1    &150.5   &87.9   &39.7    &15.0  &12.5   \\
			MWP1G024731+001580$^{1}$    & (24.73; 0.15)     &4.6    &7.3     &113.3   &198.3   &109.5  &48.6    &16.5  &6.3   \\
			\hline
		\end{longtable}}

{\scriptsize
			\begin{longtable}{lccccccccc}
			\caption{Fluxes and solid angles for the outer rings of 32 IR ring nebulae. Objects are taken from $^1$~\citet{2012MNRAS.424.2442S}, $^2$~\citet{2006ApJ...649..759C}, $^3$~\citet{2003yCat.5114....0E}.}\label{tbl-outflux}\\ \hline\hline
			Object & $l_{\rm gal}^{\circ}$; $b_{\rm gal}^{\circ}$ & $F_{8}$,Jy & $F_{24}$,Jy & $F_{70}$,Jy & $F_{160}$,Jy & $F_{250}$,Jy & $F_{350}$,Jy & $F_{500}$,Jy & $\Omega,10^{-8}$sr \\ \hline
			\endfirsthead \hline
			\multicolumn{10}{|c|}{\scriptsize\slshape(Continuation)} \\ \hline
			Object & ($l_{\rm gal}^{\circ}$; $b_{\rm gal}^{\circ}$) & $F_{8}$,Jy & $F_{24}$,Jy & $F_{70}$,Jy & $F_{160}$,Jy & $F_{250}$,Jy & $F_{350}$,Jy & $F_{500}$,Jy & $\Omega,10^{-8}$sr \\ \hline
			\endhead \hline
			\multicolumn{10}{|c|}{\scriptsize\slshape To be continued } \\ \hline
			\endfoot \hline
			\endlastfoot

			S123$^{2}$                 & (312.97; --0.43)  &120.4  &69.0   &1243.0 &3824.0 &2432.0 &1051.0 &424.2 &66.8 \\
			S167$^{2}$                 & (301.62; --0.34)  &257.3  &178.2  &1869.0 &8795.0 &6872.0 &3445.0 &1478.0&511 \\
			CN111$^{2}$                & (8.31; --0.08)    &90.5   &80.4   &1278.0 &3359.0 &1995.0 &826.8  &311.2 &31.4 \\
			N67$^{2}$                  & (35.54; 0.01)     &18.6   &20.6   &283.5  &586.0  &347.8  &155.5  &58.1  &8.12 \\
			N90$^{2}$                  & (43.77; 0.06)     &36.8   &28.1   &372.3  &1181.0 &800.5  &370.2  &148.1 &35.3 \\
			N96$^{2}$                  & (46.94; 0.37)     &2.2    &3.3    &43.1   &69.6   &42.03  &18.4   &8.5   &2.09 \\
			N98$^{2}$                  & (47.02; 0.21)     &36.3   &28.5   &403.3  &1046.0 &722.2  &331.0  &133.6 &28.7 \\
			N102$^{2}$                 & (49.69; --0.16)   &60.3   &49.8   &1015.0 &1847.0 &1134.0 &522.0  &209.2 &49.2 \\
			N121$^{2}$                 & (55.44; 0.88)     &2.4    &1.4    &12.4   &60.8   &46.0   &22.3   &9.6   &4.38 \\
			N4$^{2}$                   & (11.89; 0.74)     &173.7  &137.5  &2440.0 &4125.0 &2196.0 &914.2  &344.0 &52.4 \\
			N20$^{2}$                  & (17.91; --0.68)   &27.1   &21.7   &291.2  &901.8  &606.2  &264.0  &100.9 &14.9 \\
			N14$^{2}$                  & (14.00; --0.13)   &172.1  &265.3  &3693.0 &4792.0 &2593.0 &1019.0 &374.0 &20.1 \\
			N23$^{2}$                  & (18.67; --0.23)   &8.4    &11.5   &163.2  &295.9  &170.5  &64.8   &26.5  &2.25 \\
			N32$^{2}$                  & (23.90; 0.07)     &7.5    &8.7    &216.7  &412.5  &244.6  &103.4  &35.3  &2.17 \\
			N33$^{2}$                  & (24.21; --0.04)   &8.1    &11.3   &193.0  &382.6  &228.4  &88.1   &34.0  &2.01 \\
			N28$^{2}$                  & (21.35; --0.13)   &15.2   &12.1   &217.3  &401.2  &225.0  &86.3   &33.9  &2.95 \\
			N51$^{2}$                  & (29.15; --0.25)   &112.1  &63.4   &1082.0 &2738.0 &1629.0 &668.5  &251.5 &51.2 \\
			N60$^{2}$                  & (33.81; --0.14)   &12.5   &10.1   &203.4  &394.0  &226.6  &88.9   &37.5  &4.37 \\
			N70$^{2}$                  & (37.75; --0.11)   &7.7    &13.8   &223.3  &427.8  &269.8  &118.6  &46.0  &3.68 \\
			N80$^{2}$                  & (41.93; 0.03)     &41.9   &27.5   &370.9  &1358.0 &932.9  &431.1  &174.5 &88.5 \\
			G027.492+0.192$^{3}$       & (27.49; 0.19)     &16.3   &32.7   &526.1  &565.9  &273.2  &105.0  &33.7  &2.77 \\
			MWP1G032731+002120$^{1}$   & (32.72; 0.21)     &7.1    &62.4   &101.9  &244.1  &149.8  &64.5   &25.3  &3.55 \\
			MWP1G030250+002413$^{1}$   & (30.24; 0.24)     &15.7   &10.4   &284.1  &552.6  &296.0  &113.5  &41.9  &4.25 \\
			MWP1G017626+000493$^{1}$   & (17.62; 0.04)     &4.9    &2.9    &52.7   &176.3  &115.5  &50.2   &19.0  &2.05 \\
			MWP1G018580+003400S$^{1}$  & (18.58; 0.34)     &3.5    &3.5    &44.3   &103.1  &63.9   &27.0   &10.0  &4.67 \\
			MWP1G048422+001173$^{1}$   & (48.42; 0.11)     &6.5    &4.0    &79.6   &195.7  &122.7  &48.8   &19.1  &3.78 \\
			MWP1G024558--001329$^{1}$  & (24.55; --0.13)   &31.9   &22.4   &582.4  &1237.0 &695.8  &278.8  &110.3 &7.76 \\
			MWP1G037196--004296$^{1}$  & (37.19; --0.42)   &4.5    &5.6    &85.0   &150.9  &92.3   &40.7   &18.1  &2.14 \\
			MWP1G032057+000783$^{1}$   & (32.05; 0.07)     &18.4   &18.1   &341.8  &739.4  &457.8  &216.2  &86.0  &7.34 \\
			MWP1G018743+002521$^{1}$   & (18.74; 0.25)     &7.6    &12.5   &162.7  &276.9  &158.3  &61.4   &25.3  &2.81 \\
			MWP1G037349+006876$^{1}$   & (37.35; 0.68)     &106.9  &132.5  &196.6  &334.8  &203.2  &86.1   &34.8  &5.10 \\
			MWP1G024731+001580$^{1}$   & (24.73; 0.15)     &10.6   &14.0   &260.1  &493.4  &277.5  &109.8  &44.4  &2.57 \\
			\hline
		\end{longtable}
	}
\subsection{Spectral indices}
\begin{figure}[t]
		\centering \includegraphics[width=0.4\textwidth]{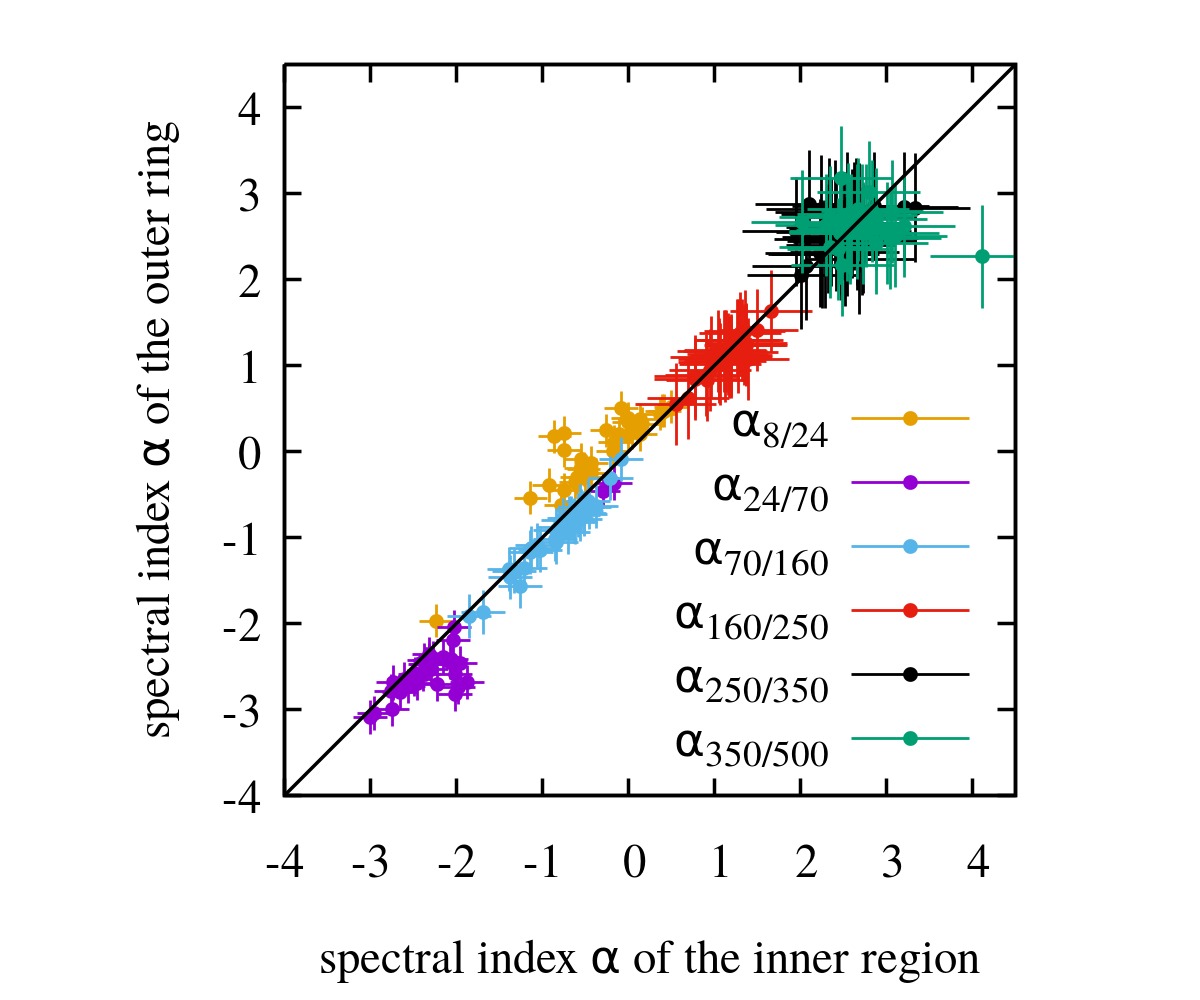}
		\caption{The relationship between the spectral indices in the inner and outer regions for the 32 IR--rings nebulae. Diagonal line represents equal spectral indices in the inner and outer parts of HII regions. Note the difference between spectral indices $\alpha_{8/24},\alpha_{24/70},\alpha_{70/160}$ in the inner and outer parts.}
		\label{fig3}
	\end{figure}

	To quantify the SED characteristic we utilize the commonly used spectral index, which can be defined as:
	\begin{equation}
	\alpha = \frac{d \log I_{\rm \nu}}{d \log \rm \nu}
	\end{equation}
	In practice, we define the spectral index for two adjacent frequencies $\nu_1$ and $\nu_2$ as $\alpha_{12}=\log{(I_1/I_2)}/\log{(\nu_1/\nu_2)}$. The comparison of the spectral indices in the inner and outer regions is presented in Fig.~\ref{fig3}.

	The spectral indices are calculated only for neighboring wavelengths as they are the most informative for our study. The shown uncertainty is calculated under the assumption of the fixed 15\% uncertainty in the flux. The diagonal line on the graph shows the position of equal indices in the inner and outer regions. One can see that the spectral indices calculated between 160 and 250~$\mu$m, 250 and 350~$\mu$m, and 350 and 500~$\mu$m ($\alpha_{160/250}, \alpha_{250/350}, \alpha_{350/500}$) are very similar in the direction to the ionized region and to the neutral shoveled envelope. This can be attributed to two reasons. First, the spatial resolution at these wavelengths may not be good enough to resolve the structure of the HII regions. Second, the emission at these wavelengths is likely dominated by the surrounding envelope, and, in the case of 3D spherical morphology, the inner aperture catches mostly emission from the front and back side of the object and not from the ionized vicinity of the central star(s).

	The spectral indices $\alpha_{8/24}, \alpha_{24/70}, \alpha_{70/160}$ are clearly different in the inner and outer parts. If the points in Fig.~\ref{fig3} lie above the diagonal line that means that the ratio of the shorter-wavelength intensity to longer-wavelength intensity is higher in the outer region. This case is represented by the slope between 8 and 24~$\mu$m, $\alpha_{8/24}$, and is the simple consequence of the ring-like morphology at 8~$\mu$m while intensities at 24~$\mu$m are comparable in the inner and outer regions. This leads to the smaller difference between intensities at 8 and 24~$\mu$m in the outer region than in the inner region. As the 8~$\mu$m emission is dominated by PAHs and 24~$\mu$m emission comes mostly from the VSGs, the slope $\alpha_{8/24}$ is probably uninformative of the thermal structure, but rather reflect the density distribution of PAHs and VSGs.

	The spectral indices $\alpha_{24/70}, \alpha_{70/160}$ are definitely larger within the inner aperture than in the outer aperture (check magenta and blue dots on Fig.~\ref{fig3}). This is exactly what expected for the object which is hotter inside and colder outside. The 24~$\mu$m emission is stochastically generated, which makes it harder to use as a tracer of the temperature. On the other hand, the emission at $70-160~\mu$m is likely {\it not} stochastic and comes from the dust with relatively similar size distribution in the inner and outer regions~\citep{Akimkin_17}. This makes the color $\alpha_{70/160}$ the best proxy of dust temperature in HII regions.

\begin{figure}[t]
		\centering \includegraphics[width=0.4\textwidth]{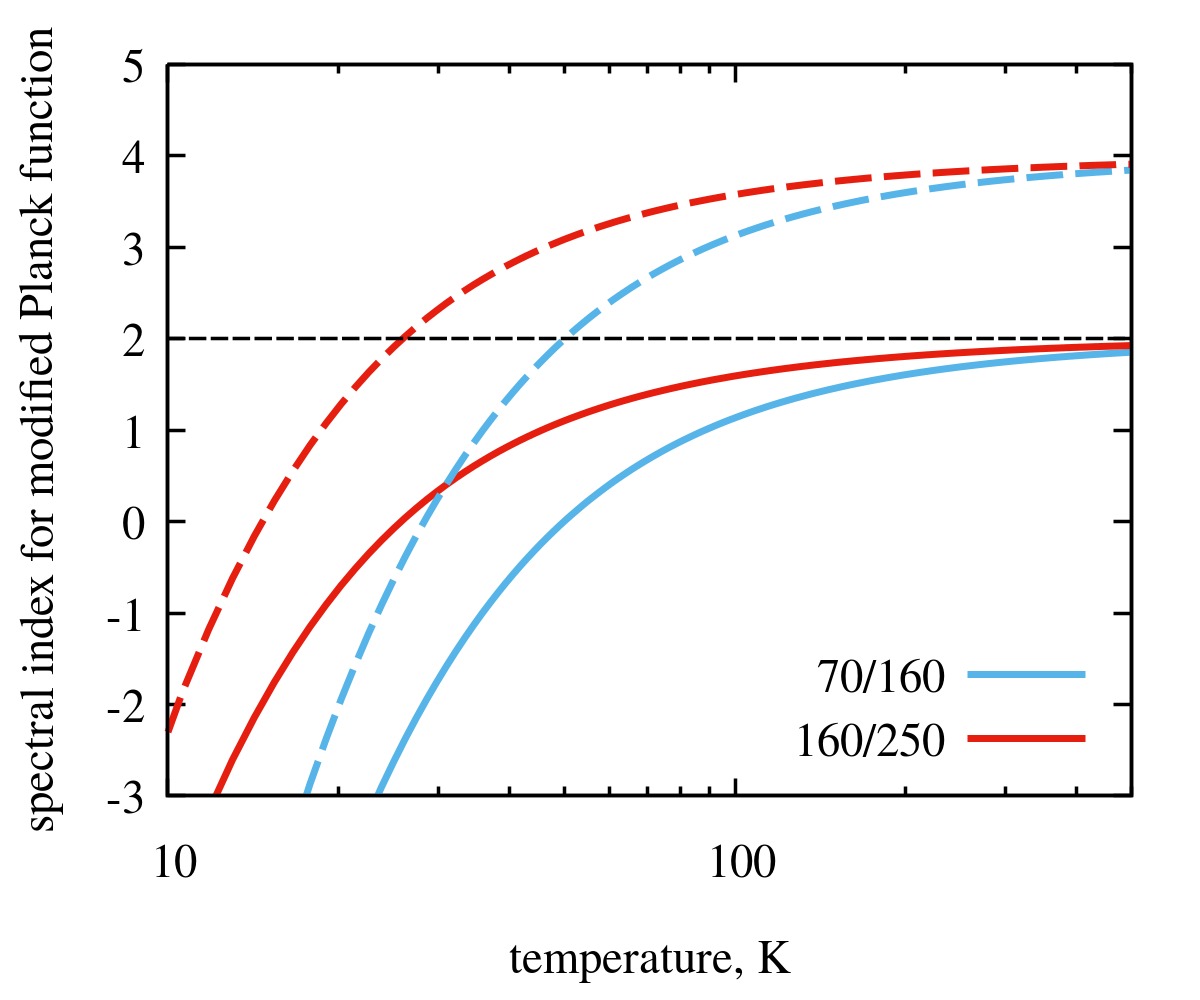}
		\caption{The spectral index $\alpha$ between 70 and 160~$\mu$m (blue) and 160 and 250~$\mu$m (red) for the modified blackbody emission. Solid and dashed lines show optically thick and optically thin limits, respectively.}
		\label{fig4}
	\end{figure}

	For the illustration of the spectral index dependence on the dust temperature, we calculate it for the case of the modified blackbody emission $I_{\nu}=(1-e^{-\tau_{\nu}})B_{\nu}(T)$. Here $B_{\nu}(T)$ is the Planck function for a given dust temperature, and $\tau_{\nu}$ is the optical depth. The corresponding spectral index is
	\begin{equation}
	 \alpha=3-\frac{x}{1-e^{-x}}+\frac{\tau_{\nu}}{e^{\tau_{\nu}}-1}\beta,
	\end{equation}
	where $x=h\nu/(k_BT)$ and $\beta= d \log \varkappa_{\rm \nu}/ d \log \rm \nu$ is the opacity index characterizing the spectral properties of the dust opacity coefficient $\varkappa_{\nu}$, cm$^2$ g$^{-1}$.
	To calculate $\varkappa_{\nu}$ we use optical properties of silicate and carbonaceous grains as in \citet{Akimkin_17}, but assuming power-law size distribution with the slope $-3.5$, grain sizes between $0.005$ and $0.25\mu$m and mass ratio 0.2:0.8 between carbonaceous and silicate dust. The corresponding spectral indices $\alpha_{70/160}$ and $\alpha_{160/250}$ for the cases of optically thin and thick emission is presented in Fig.~\ref{fig4}).
	The horizontal dashed line shows the Rayleigh-Jeans limit $\alpha=2$ in optically thick media (the corresponding limit in the optically thin case is $\alpha=2+\beta$). Note that $\alpha$ monotonically increases with temperature. The observed values of $\alpha_{70/160}$ lie between $-2$ and 0, which corresponds to a temperature range of $20-30$\, K. While the inner values of $\alpha_{70/160}$ are definitely larger than the outer ones, indicating the hotter interiors, the difference between them is quite small and translates to a corresponding difference in temperatures smaller than several kelvins. One may expect a much larger difference considering the temperature gradient between the ionized region and neutral envelope. This is again indicative of 3D morphology and a corresponding significant contribution from the envelope to the radiation in the inner aperture. For the restoring of the dust properties inside HII regions more detailed approach is needed.

	\subsection{Signs of dust evolution processes}

	One of our aims is to make a step towards the disentangling the possible mechanisms removing dust from the HII regions. The photo-destruction is more efficient around more energetic stars, while dust removal via radiative drift is on the opposite. The balance between the radiation pressure and the friction of charged dust with plasma is in favor of the latter around the stars with higher effective temperatures~\citep{Akimkin_17}. Thus, in the pure drift (no destruction) model one may expect more dust around more energetic stars. On top of that, both mechanisms should operate with different efficiency for different grain sizes as photo-destruction is necessary to explain the ring emission at 8~$\mu$m~\citep{2013ARep...57..573P}.

\begin{figure}[t]
		\centering
		\includegraphics[width=0.3\textwidth]{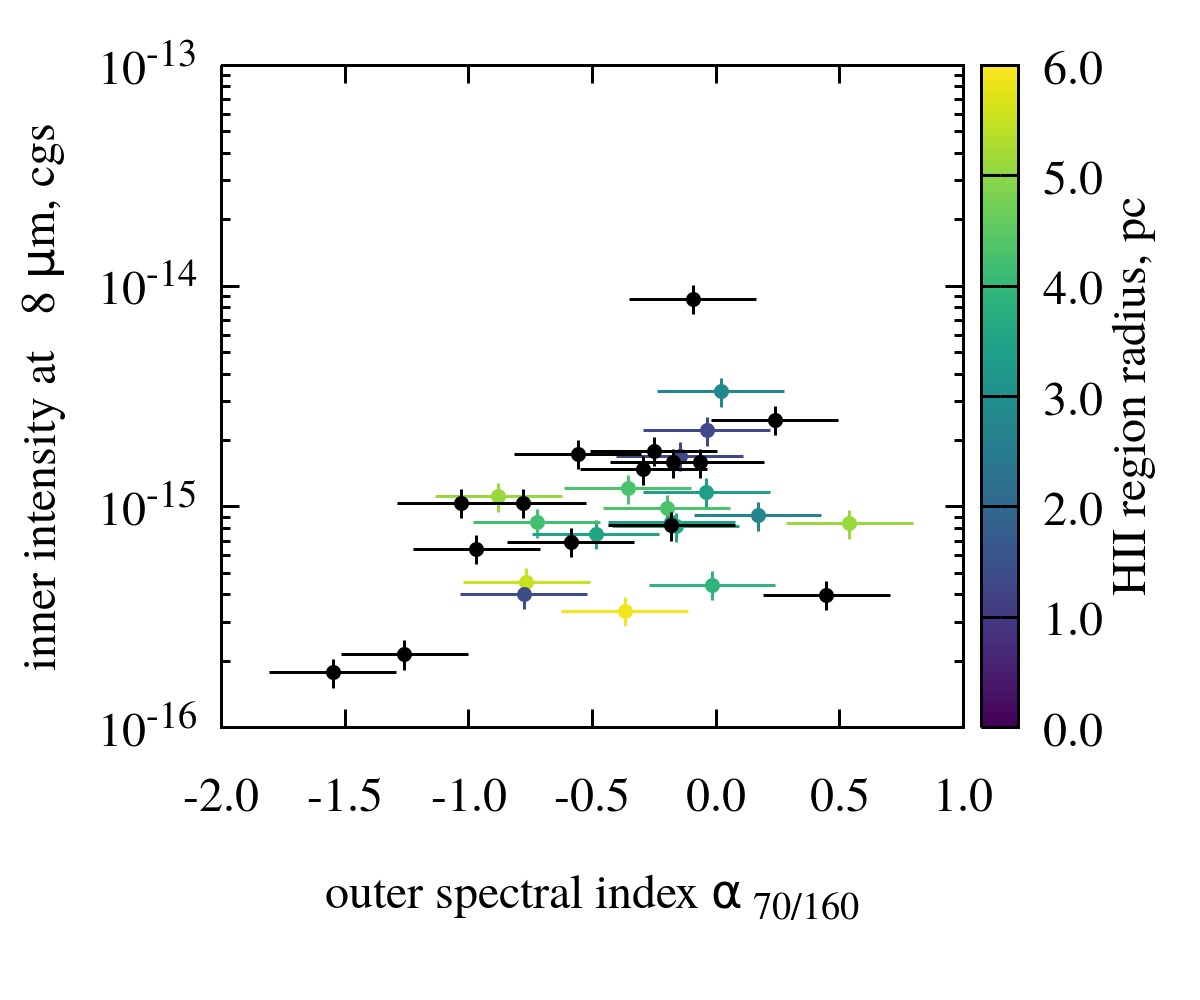}
		\includegraphics[width=0.3\textwidth]{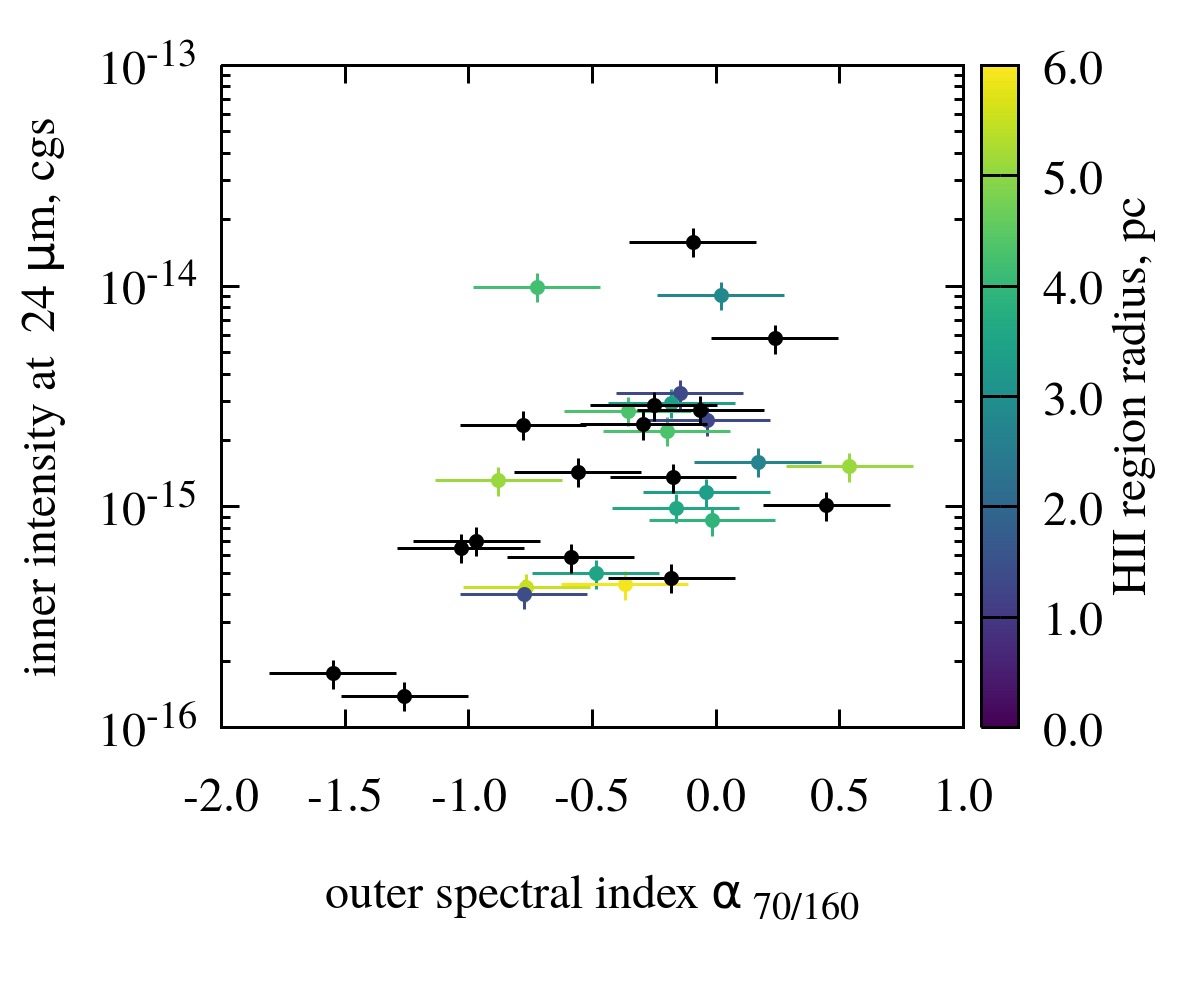}
		\includegraphics[width=0.3\textwidth]{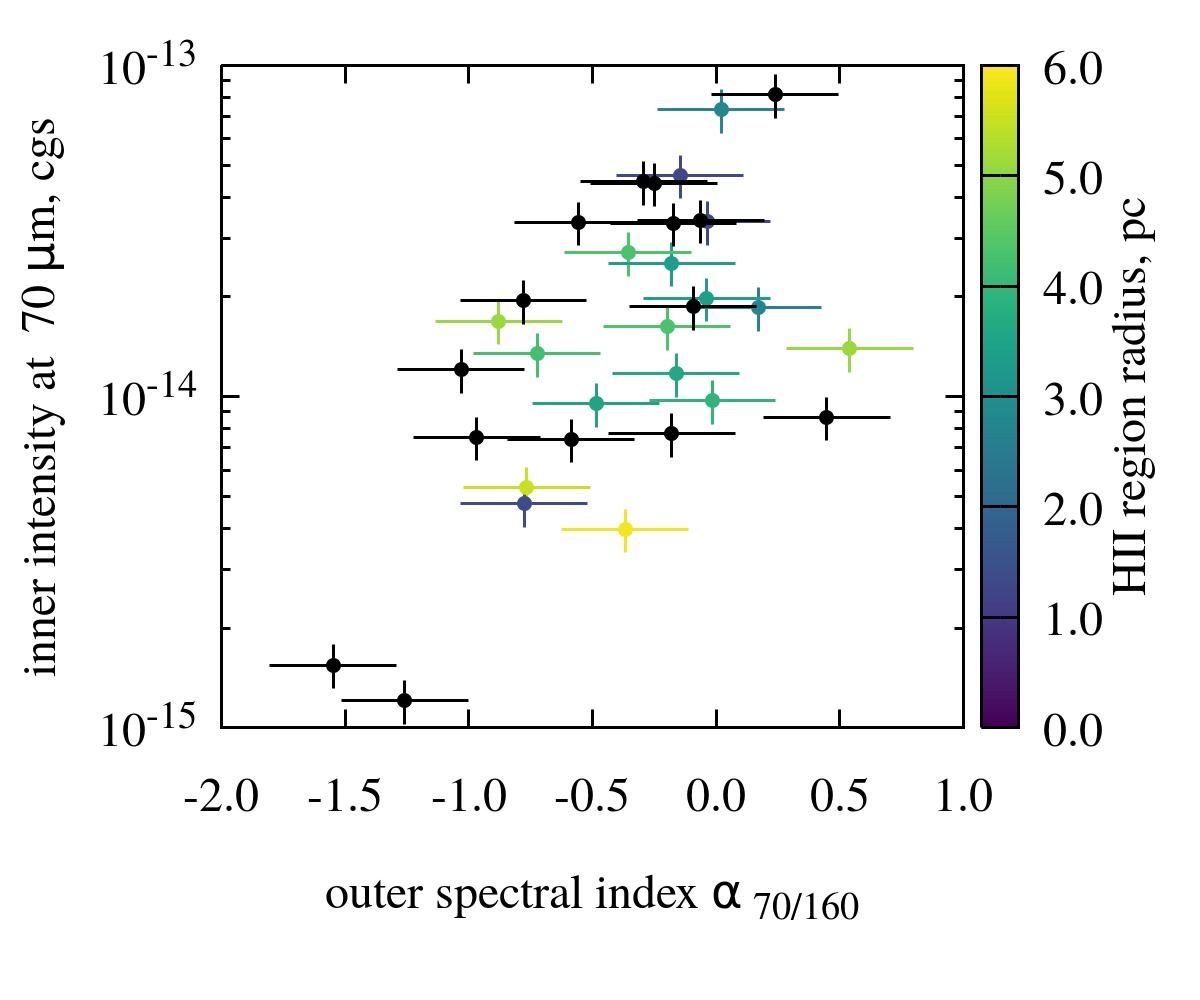}
		\caption{Average inner intensity at 8, 24, and 70~$\mu$m as a function of the spectral index $\alpha_{70/160}$ in the outer region. Dot color denotes the linear size of the ionized region (black if not known).}
		\label{fig5}
	\end{figure}

	As the color index $\alpha_{70/160}$ positively correlates with dust temperature and is not very sensitive to the amount of dust in the optically thin regime, it can be used as an indicator of the envelope temperature. The envelope temperature depends on the distance to the ionizing star and stellar effective temperature. The former can be accounted if the distance to the source is known, which is true for 17 of 32 our sources. Thus one may consider $\alpha_{70/160}$ as a rough proxy of the stellar effective temperature. The average intensity inside the inner aperture depends on both temperature and amount of matter in the direction to the central ionizing source. The critical question here is the proportion between surface densities of dust in the envelope and in the HII region itself. In the case of 3D morphology of IR bubbles, the amount of dust in the neutral envelope is likely sufficiently higher than in the inner ionized region, while 2D toroidal morphology implies that the dust emitting  from the inner aperture is actually located inside the ionized region. In Fig.~\ref{fig5} we show the dependence of the average intensity within the inner aperture at 8, 24, and 70~$\mu$m on the outer spectral index $\alpha_{70/160}$.
	In the case of 2D ring morphology (no contribution of the envelope to the inner aperture) and strong photo-destruction of PAHs and other grains emitting at $8-70\,\mu$m one may expect an anti-correlation between the presented parameters. Obviously, this is not what is observed, leaving us with several options. First, the increase in the inner intensity with $\alpha_{70/160}$ can be simply explained by the higher temperature envelope contributing to the intensity towards the ionized region. Second, the observed behavior may imply the better retention of dust around more energetic stars. Third, a moderate photo-destruction may be veiled by the stronger emission of grains surviving the destruction.

\begin{figure}[t]
		\centering
		\includegraphics[width=0.3\textwidth]{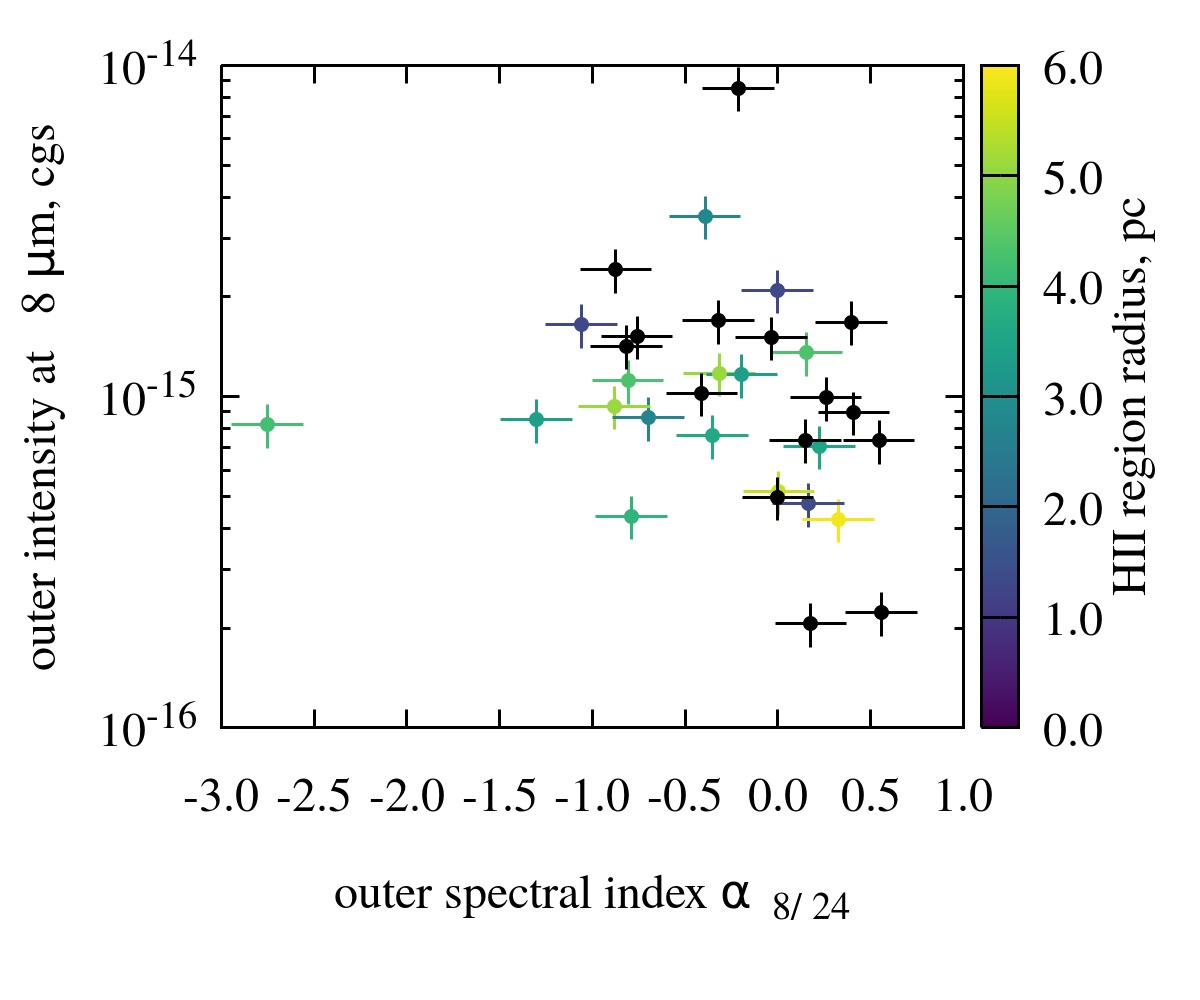}
		\includegraphics[width=0.3\textwidth]{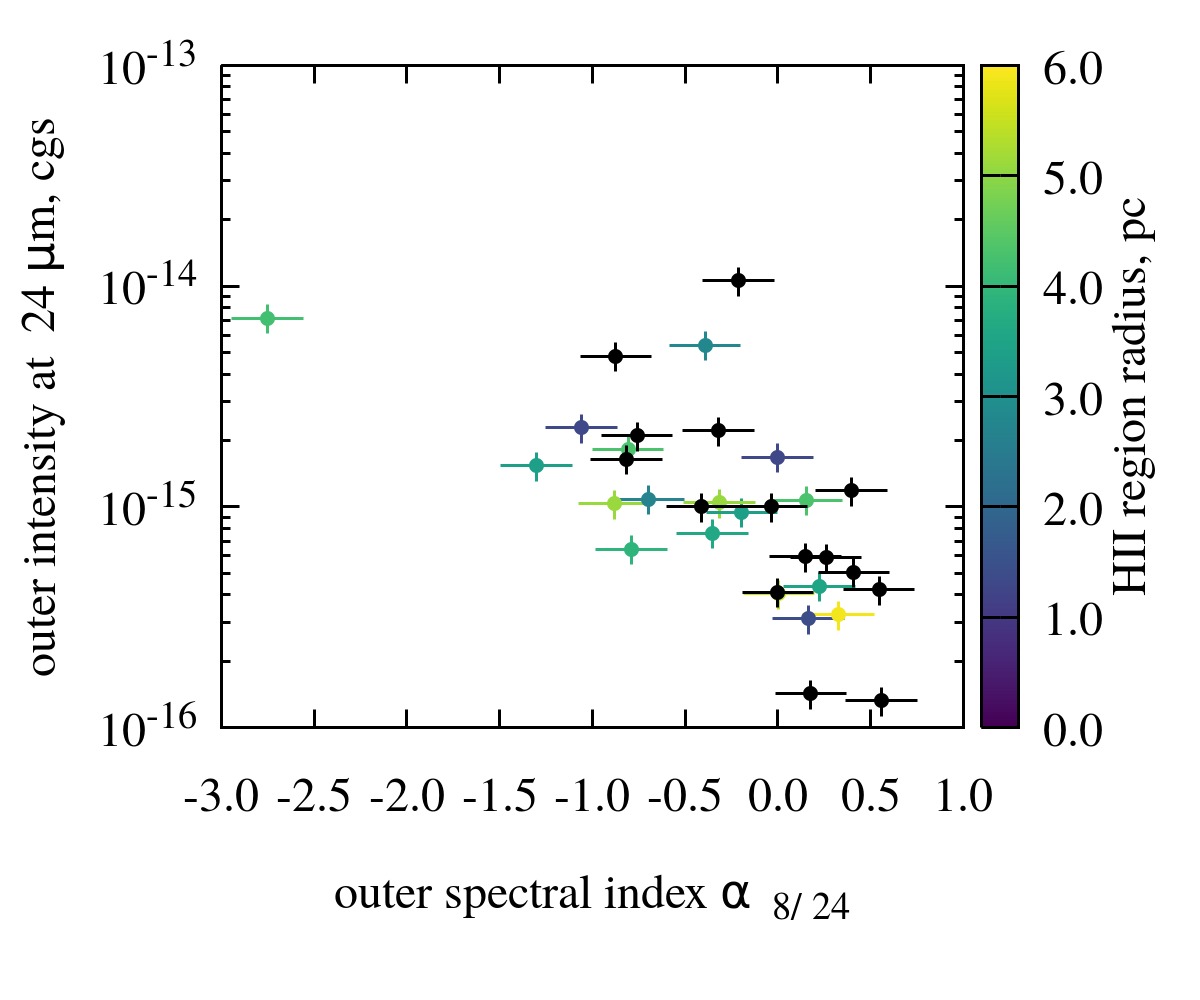}
		\includegraphics[width=0.3\textwidth]{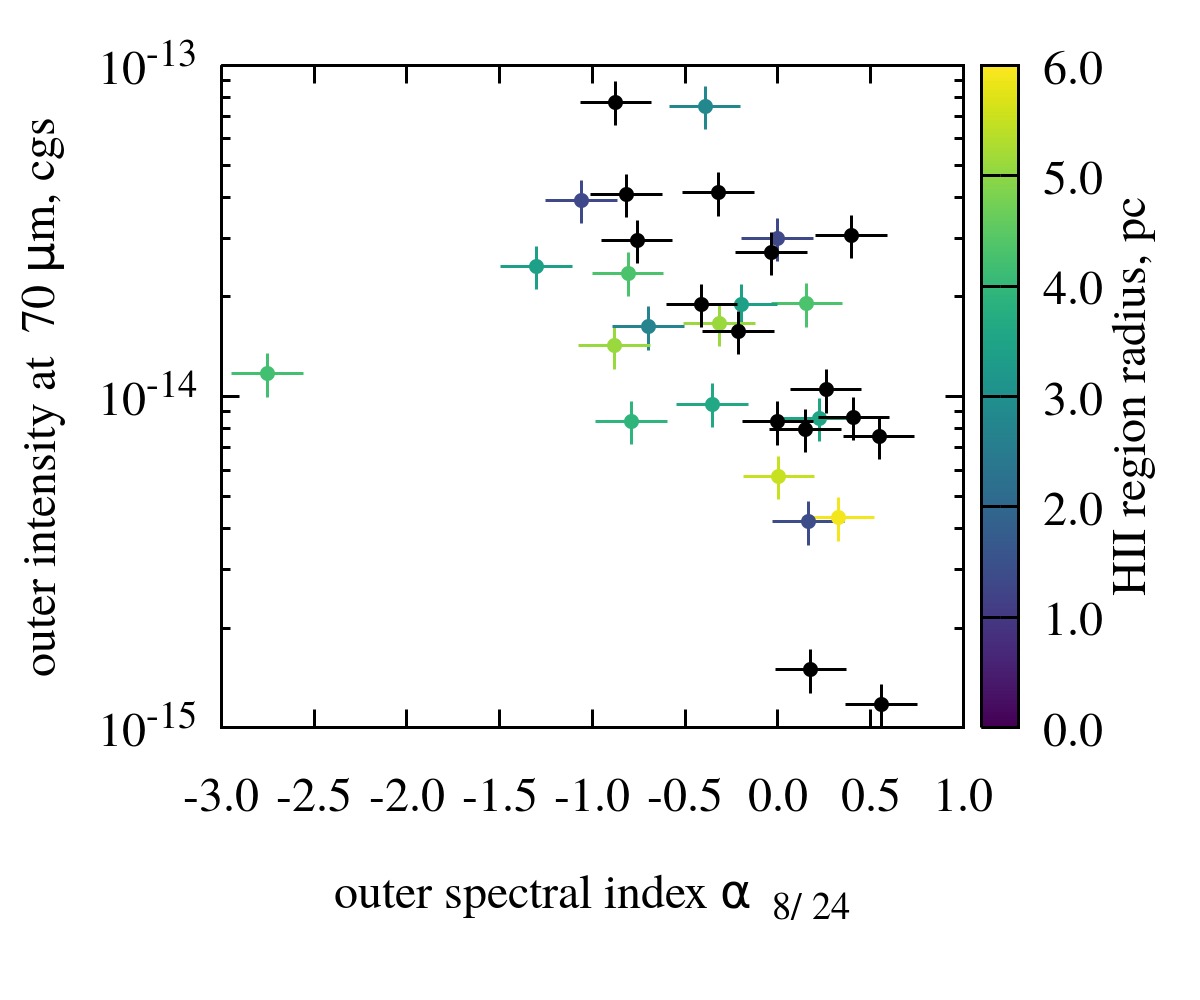}
		\caption{Average outer intensity at 8, 24, and 70~$\mu$m as a function of the spectral index $\alpha_{8/24}$. Dot color denotes the linear size of the ionized region (black if not known).}
		\label{fig6}
	\end{figure}

	The other interesting relation is the map $\alpha_{8/24}$ vs. intensity at 8, 24, and 70~$\mu$m in the outer regions (Fig.~\ref{fig6}). The variation of $\alpha_{8/24}$ can be primarily caused by variations in the intensity at 8~$\mu$m, 24~$\mu$m, or both. Theoretically, if larger values of $\alpha_{8/24}$ were determined solely by the increase in 8~$\mu$m intensity with relatively constant 24~$\mu$m intensity, there would be a positive correlation between $\alpha_{8/24}$ and intensity at 8~$\mu$m. In the second case (constant 8~$\mu$m, varying 24~$\mu$m), the anti-correlation between $\alpha_{8/24}$ and intensity at 24~$\mu$m would be expected. In the third case (both intensities are varying), no clear trends should be present. As can be seen from the first and second panels of Fig.~\ref{fig6}, the second case is more likely, i.e. the variations of $\alpha_{8/24}$ are caused mostly by the variation in 24~$\mu$m emission. The slope $\alpha_{8/24}$ can trace the relative abundance between PAHs and VSGs (with larger values of $\alpha_{8/24}$ corresponding to the larger PAHs fraction). The intensity at 70~$\mu$m depends on both the surface density of dust and its temperature. Thus, the distribution as seen on the third panel of Fig.~\ref{fig6} may be interpreted as the observational evidence of the transformation of large dust into PAHs. We plan to investigate the validity of this hypothesis using theoretical modeling with MARION and SHIVA tools~\citep{Akimkin_17, Murga_16} in future studies.

	\section{Conclusions}

	We present an extension of the IR ring nebulae catalog by~\citet{Topchieva_2017} based on {\it Spitzer} and {\it Herschel} data. We selected 32 of 99 objects with the most regular rounded shapes and calculated fluxes separately from their inner parts and outer rings at seven wavelengths from 8 to 500~$\mu$m. The outer rings correspond to the neutral shoveled envelopes, while inner apertures may catch both the radiation from ionized parts of the objects and from the front and back walls of the neutral envelopes. From the analysis of the spectral slopes in the inner and outer SEDs of the whole ensemble we found that:
	\begin{itemize}
		\item the spectral index $\alpha_{70/160}$ between 70 and 160~$\mu$m is the most appropriate tracer of dust temperature in the envelope. The use of shorter wavelengths is complicated by the stochastic nature of the emission by PAHs and VSGs, while the longer wavelengths suffer from the insufficient spatial resolution hampering the differentiation between inner and outer parts of HII regions. More importantly, the emission from dust heated to the envelope temperatures ($20-30$\,K) lies between 70 and 160~$\mu$m, which ensures good sensitivity of $\alpha_{70/160}$ to the dust temperature;
		\item the difference between the spectral indices $\alpha_{24/70}$ and $\alpha_{70/160}$ of the inner and outer regions does indicate of hotter interiors, but is too small in the view of expected temperature difference of ionized and neutrals parts of the regions. This indicates that the inner aperture is significantly contaminated by the envelope radiation, i.e. hints on 3D rather than 2D morphology of IR bubbles.
		\item objects with hotter envelopes (traced by larger $\alpha_{70/160}$) are more likely to show higher radiation intensity from the inner parts at 8, 24, and 70~$\mu$m. Again, this is a signature of a heated spherical envelope (hotter dust produces more radiation) and unfortunately does not allow to differentiate between radiation pressure and photo-destruction as dominant mechanism removing dust from ionized regions. (More energetic stars favor better photo-destruction and less effective radiation drift). However, 2D ring-like morphology with dominating photo-destruction can be excluded based on these observations.
		\item the prevalence of PAHs over very small grains (if traced by $\alpha_{8/24}$) is accompanied with smaller fluxes at 24 and 70~$\mu$m. This may provide an observational indication that PAHs are formed due to the destruction of larger grains.
	\end{itemize}
	The proper modeling tackling dust destruction and dynamics in expanding HII regions is needed to support these conclusions from the theoretical point of view~\citep{Murga_16, Akimkin_17}. This could allow to determine how different dust evolution processes (photo-destruction, radiative drift, aliphatisation, etc) manifest themselves in the IR synthetic images.

	\normalem
	\begin{acknowledgements}
		The authors are grateful to Ya.~N.~Pavlyuchenkov, D.~S.~Wiebe, M.~S.~Kirsanova for useful comments. We also thank the referee for a thorough and careful
reading of the manuscript. The work was supported by a grant from the Russian Foundation for Basic Research 18-32-00384.

		We also thank the developers of numpy~\citep{numpy}, scipy~\citep{scipy}, and matplotlib~\citep{matplotlib} Python packages.

	\end{acknowledgements}

	\bibliographystyle{raa}
	\bibliography{msRAA_2019_0068_Topchieva_AP_2019}

\end{document}